\documentclass[runningheads]{svmult}

\usepackage{makeidx}   % allows index generation
\usepackage{graphicx}  % standard LaTeX graphics tool
\usepackage{subeqnar}  % subnumbers individual equations
\usepackage{multicol}  % used for the two-column index
\usepackage{physprbb}  % modified textarea for proceedings,
\makeindex             % used for the subject index
\newcommand{\greeksym}[1]{{\usefont{U}{psy}{m}{n}#1}}
\newcommand{\umu}{\mbox{\greeksym{m}}}

\def \ion#1#2{#1~{\small\rm{#2}}\relax}
\def \HI       {\ion{H}{I}}
\def \HII      {\ion{H}{II}}
\def \NII      {[\ion{N}{II}]}
\def \OII      {[\ion{O}{II}]}
\def \OIII     {[\ion{O}{III}]}
\def \ha       {H$\alpha$}
\def \hb       {H$\beta$}
\newcommand{\eg}{e.g.~}

\newcommand{\magarc}{\ifmmode {{{{\rm mag}~{\rm arcsec}}^{-2}}}
             \else {{{mag}$~${arcsec}$^{-2}$}}
             \fi}
\newcommand{\kms}{\ifmmode\,{\rm km}\,{\rm s}^{-1}\else km$\,$s$^{-1}$\fi}

\def \famag{\hbox{$.\!\!^{\scriptsize m}$}}

\def\gtsim{\mathrel{\vcenter{\offinterlineskip \hbox{$>$}
    \kern 0.3ex \hbox{$\sim$}}}}
\def\ltsim{\mathrel{\vcenter{\offinterlineskip \hbox{$<$}
    \kern 0.3ex \hbox{$\sim$}}}}
\begin{document}
\title*{Galaxy Evolution from Emission Linewidths}
\toctitle{Galaxy Evolution from Emission Linewidths}
% allows explicit linebreak for the table of content
%
\titlerunning{Galaxy Evolution from Emission Linewidths}
\author{St\'ephane Courteau\inst{1}
%\and Sandra Faber \inst{2}
\and Young-Jong Sohn \inst{2}}
\authorrunning{Courteau \& Sohn}

\institute{Univ. of British Columbia, Dept. of Physics \& Astronomy, BC 
  V6T 1Z1 \ Canada
% \and University of California, UCO/Lick and Dept. of Astronomy and 
%      Astrophysics, Santa Cruz, CA 95064, USA
\and Yonsei University, Center for Space Astrophysics, Seoul 120-749, Korea}

\maketitle              % typesets the title of the contribution

\begin{abstract}

The major thrust of the Tully-Fisher (TF) surveys of distant galaxies 
is the measurement of {\it linewidths} rather than mere redshifts 
or colors. Linewidths are a measure of galaxy mass and should 
therefore be a more stable indicator of size than galaxy brightness, 
which can be badly affected by luminosity evolution.  Masses may provide 
the best way to relate galaxies at different epochs, but for such a 
program to work, we must control systematic effects that could bias 
linewidth measurements at high redshift and skew comparisons with 
local Tully-Fisher calibrations.  Potential sources of confusion 
in TF studies of galaxy structure and evolution include central or
extended star bursts, infalling gas, turbulence and outflows, dust 
extinction, calibration of emission linewidths, and improper application
of local TF calibrations to high redshift galaxies.

% With these effects accounted for, we foresee an exciting future 
% for TF studies of galaxy formation and evolution. 

\end{abstract}

\section{Introduction}

Studies of galaxy structure and measurements of galaxy distances have
often relied on the tight match between absolute luminosity and rotational 
linewidth, or Tully-Fisher relation (TFR).  The best method for measuring 
spiral linewidths {\it locally} ($z \leq 0.05$) uses \ha\ rotation 
curves.  For distant galaxies with $z > 0.4$, 
H$\alpha$ is shifted out of the optical bandpass into a forest of 
night-sky lines, forcing reliance on bluer emission lines 
(\OII$\lambda$3727, \hb\ and \OIII$\lambda\lambda$4959,5007).
Blue profiles may be affected by dust absorption more 
severely than \ha, and the radial distribution of \OII\ and \OIII\ 
will depend on the abundance and temperature of the \HII\ regions in a 
complicated way compared to \ha.  For intermediate redshifts,
spatially-resolved kinematics are no longer possible.  Thus,
one must first establish a calibration of blue and red, resolved[2D]
and integrated [1D], galaxy linewidths in order to tie the local TFR 
to distant galaxies. 

Courteau (1992; 1997; hereafter C92, C97) demonstrated the tight correlation 
between optical rotation curves[2D] and profiles[1D] and single-dish 
21~cm profiles for a large sample of late-type spirals (see also 
Mathewson et~al. 1992).  Kobulnicky \& Gebhardt (2000) extended this 
correlation to \OII\ emission lines using a broad sample of Hubble 
types and showed that reliable linewidths
%, and thus galaxy masses,  %% Not obvious - not virial masses...
could be obtained for distant galaxies.  Blue linewidths for distant 
galaxies can therefore be compared to a local foil (\ha\ or \HI) without 
introducing additional systematic effects. We confirm, and expand upon,
the results of Kobulnicky \& Gebhardt (2000) with a more extensive
sample of Sc galaxies below.  We also caution about important caveats in 
the application and interpretation of the TFR for distant galaxies. 

\subsection{Calibration of Blue Linewidths}

Courteau (and Sandra Faber) collected optical long-slit spectra 
(\OII, \hb, \OIII, \ha, \NII) of 20 Sc galaxies with the Lick 3-m 
telescope and Kast Double spectrograph (Courteau, Faber, \& Sohn 2002;
hereafter CFS02).  
Integration times were 1800s for each spectrum and R$\simeq6000$.
Our observing strategy includes the measurement of narrow-slit major
axis spectra and drift scans over the full size of the galaxy to
simulate low-resolution integrated profiles as measured at large distances.
% Our observing strategy was divided into 3 parts: 1) Collect blue spectra for 
% the whole sample using a narrow slit to compare with 
% best-resolution \ha\ and \HI\ measurements; 2) Separately study the effects 
% of decreased spatial resolution on rotation curves and linewidths to mimic 
% integrated profiles as seen at large distances. This is done with a 
% drift-scan technique along the minor-axis of a subset of galaxies; 
% by varying the length of a scan, we can simulate distance effects.
% We take two scans per galaxy covering half or all of a galaxy's projected 
% surface on the sky; 3) Extract similar scans from the same subset as in (2) 
% but at H$\alpha$. 
The data were reduced according to standard procedures (C97); 
rotation curves were extracted by measuring intensity-weighted centroids 
at each spatial bins and accounting for instrumental broadening.  The \OII\ 
doublets were fitted with a double Gaussian function convolved with the 
instrumental profile. 
The extracted rotation curves were then modelled with a smooth function,
e.g. an arctan, to obtain characteristic terminal or suitably 
chosen velocities.  For late-type spirals, the velocity estimate, $V_{2.2}$, 
at 2.2 disk scale lengths (or equivalently $V_{1.3}$ at 1.3 effective radius)
minimizes TF scatter (C97, Courteau \& Rix 1999).
%; see also Kannappan et~al. 2002). 
%a measure of the terminal 
%velocity is also required for the TF relations of low-surface brightness 
%and high-surface brightness galaxies to match simultaneously 
%(Zwaan et~al. 1995).  

\smallskip

\begin{figure}[th]
\begin{center}
\includegraphics[width=0.89\textwidth]{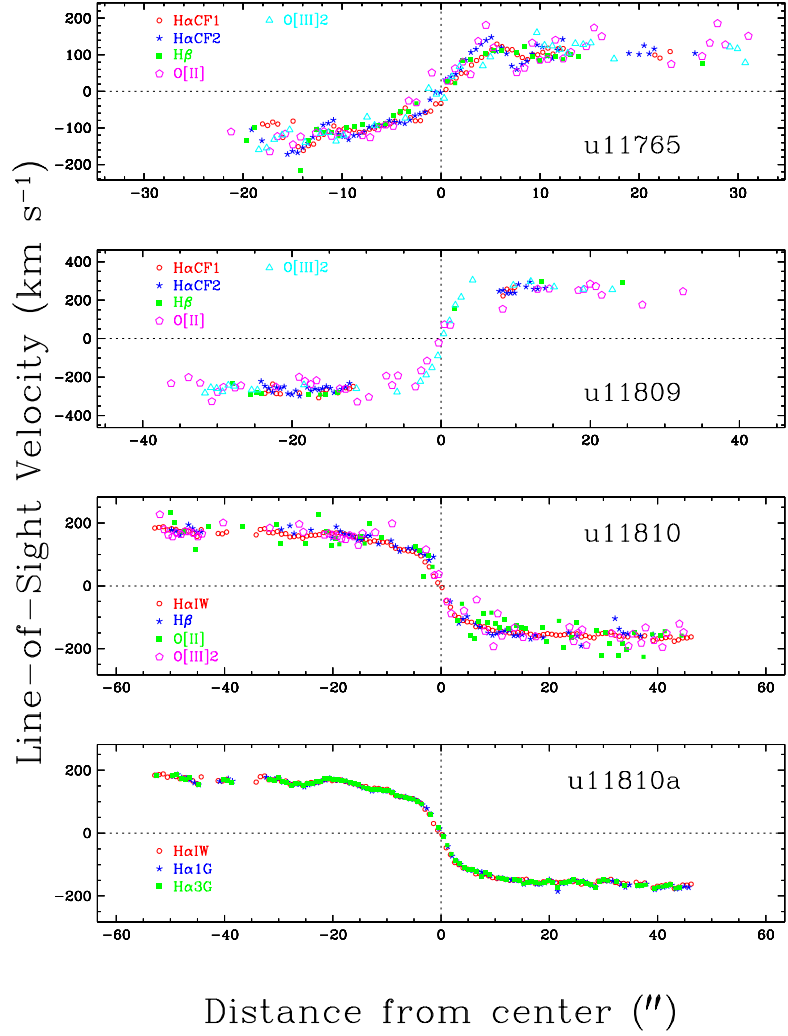}
\end{center}
\caption[]{Example of optical rotation curves for 3 UGC late-type spiral
galaxies. H$\alpha$CF refers to the compilation of \ha\ linewidths by 
Courteau (1992).
The notation H$\alpha$IW, H$\alpha$1G, and H$\alpha$3G refers to velocity
centroids extracted using an intensity-weighted scheme, single-Gaussian 
fits (to the \ha\ line), and triple-Gaussian fits (to the \ha-\NII\ complex).
The different fitting techniques yield comparable results, as seen in the 
bottom panel.  The match between different kinematic tracers (top three
panels) is excellent.}
\label{RCs}
\end{figure}

\begin{figure}[th]
\begin{center}
\includegraphics[width=0.7\textwidth]{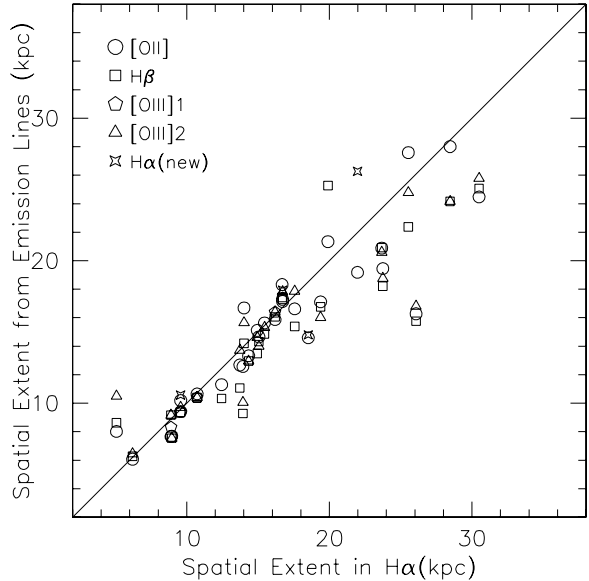}
\end{center}
\caption[]{Maximal extent of optical emission lines in late-type spirals.
 The \ha\ measurements (x-axis) are taken from the CF sample (C92, C97).}
\label{RCext}
\end{figure}

We verify that blue and red emission lines map the same velocity 
field (\eg\ Fig.~\ref{RCs}) though differences in spatial coverage,
or flux ratios, exist.  Thus, smooth fits to resolved rotation curves 
should yield matching $\Delta V$'s at all wavelengths.  The rotation 
curve for UGC 11809 (Fig.~\ref{RCs}) illustrates significant emission
differences for \ha\ and blue linewidths at the center and outskirts 
of the galaxy.  On average, blue and red emission lines trace the
same extent (see Fig.~\ref{RCext}), but for large systems ($R_{\rm max} >
20$ kpc), \ha\ is systematically detected beyond the last point of
blue emission. 
%(agrees with Kennicutt, Bland-Hawthorne ...)

\smallskip
We also construct 1-dimensional emission line profiles by collapsing the 
2-dimensional rotation curves along the spatial axis (We account for 
the varying fraction of light at each radii covered by the long slit.)
These profiles mimic the integrated linewidths of distant galaxies, and
can be compared with lower resolution (and more realistic) drift scans. 
A suitable definition of integrated linewidth,
measured at 20\% of total flux, yields a good match to local 21cm 
linewidths (C97), both for collapsed profiles and drift scans.  Note 
that it is {\it critical} that linewidths be measured similarly for 
calibrators and targets to prevent artificial biases.   

\smallskip
We find a good correlation, with 15-20\% scatter, between \OII, \hb, 
and \ha\ {\it integrated} linewidths (CFS02).  
\OIII\ fluxes are comparatively 
low and [OIII] integrated profiles are generally too noisy for
extragalactic investigations.  For emission-line galaxies, the integrated 
flux at [OII] is on average four times greater than at \hb; \OII\ might 
therefore be favored as a kinematic tracer at high redshift.  
\hb\ $\Delta V$'s are however more easily measured (the \OII\ doublet 
requires careful deblending). 
% \footnote{ Kobulnicky \& Gebhardt (2000) did not measure \hb\
% emission linewidths.}.
% Kennicutt (1992) suggests that \hb is not as reliable a tracer 
% of star formation due to variations in stellar absorption and excitation.  

\smallskip
Also, in contrast with Kennicutt (1992), Jansen et~al. (2001) show 
that \hb\ is a considerably better tracer of star formation than \OII. 
We suggest that \hb\ and \OII\ emission lines be secured whenever
possible for linewidth confirmation and star formation estimates 
(see e.g. Charlot \& Longhetti 2001). 

\newpage

% Figure 1

% Figure 2

\subsection{Central Star Bursts}

At intermediate redshifts, the rotation curves of spiral disks can no 
longer be resolved and studies of TF evolution must rely on integrated 
linewidths. 

The major drawback of intensity-weighted profiles is that a flux
integral favors the brightest \HII\ and star-bursting regions (C92, C97). 
Unlike resolved rotation curves, integrated linewidths also suffer from
seeing and instrumental broadening effects which further complicates 
their analysis. 

Central bursts may cause important linewidth (mass) discrepancies 
beyond $z \sim 1$ where the population of bursting galaxies 
increases significantly (e.g. Steinmetz; this conference [tc]). 
Linewidths are usually measured relative to peak or total fluxes; 
either way, the effect of {\it central} starburst activity biases 
linewidths {\it low}.  Galaxies with central bursts thus appear too 
bright (in a TF sense) for their linewidths, thereby mimicing luminosity 
evolution.  

%Figure 3 
\begin{figure}[th]
\begin{center}
\includegraphics[width=0.7\textwidth]{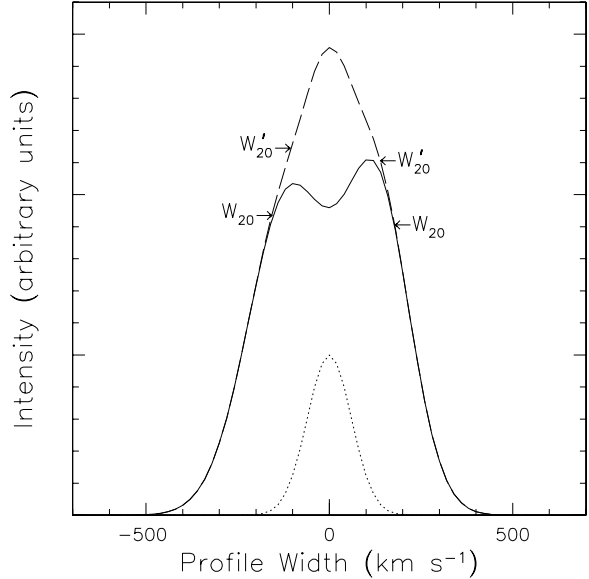}
\end{center}
\caption[]{Schematic representation of intensity-weighted rotation profiles 
  with (dashed curve) and without (solid curve) a central starburst,  
  based on an average of observations (C92). 
 Linewidths measured at a fraction (e.g. 20\%) of the total flux
 are biased {\it low} by the effect of a central starburst.  The 20\% 
 linewidth baseline with and without starburst is measured between 
 W$^\prime_{20}$ and W$_{20}$, respectively.  The broad Gaussian-like
 emission profile from the star-bursting bulge is also shown as
 a dotted curve.}
\label{burst}
\end{figure}
 
C92 compared linewidths of nearby galaxies with central
starburst/AGN activity from resolved and integrated rotation profiles 
(see Fig.~\ref{burst}) and found that star-bursting bulges can 
yield linewidth measures that are 25-30\% smaller than the true 
dynamical value.  This corresponds to artificial offsets of $\sim$
0\famag8 from the nominal TF relation, or a 75\% increase in luminosity.  
Similarly, half-light radii 
%(concentration indices) can also be reduced (increased) 
can also be reduced by 40\% or more by starburst activity.  This
{\it ``starburst bias''} may explain the preponderance of blue 
compact (bursting) objects at intermediate redshifts (e.g. Guzman et~al. 
1997; Hammer et~al. 2001; Koo [tc]) which would be the centers of massive, 
$L^*$ or less, spiral galaxies (see e.g. Barton \& van~Zee 2001 [BvZ01]). 
BvZ01 have also considered the effect of star-bursting bulges on integrated 
linewidths and find somewhat milder trends than reported here but do not 
preclude more active bursts and thus larger offsets such as the ones we find. 
Effects which would bias integrated linewidths {\it high} include extended 
(non-central) star-bursts, infalling gas from satellites and dwarfs, 
turbulence, gas outflows, etc.  Alternate definitions of linewidth 
measurements may help control some of these biases (CFS02). 

AGN activity, from mergers or interactions, is likely to 
increase with redshift at a rate $\propto z^{2.3}$ (Patton et~al. 2002), 
% The considerations above may thus be viewed as casting gloom on
% kinematical studies of distant galaxies.   
but we may be 
able to weed out bursting systems on the basis of their blue
colors, especially if they are nuclear AGNs rather than starbursts
(Kannappen et~al. 2002), to the extent that a color-TF residual 
correlation can be identified at high $z$.  Furthermore, little 
luminosity evolution has so far been detected up to $z\sim1$ 
(e.g. Koo [tc]) and recent studies of TF evolution at intermediate 
redshifts based on integrated linewidths suggest only minor, if any, 
effects of central bursts (Vogt et~al. 1997; Weiner [tc]).

\newpage

\subsection{Dust Extinction}

The likely higher fraction of dust in the central parts of spiral galaxies 
at high redshifts is also a cause for concern.  Dust extinction in the 
inner disk steepens the solid body part of the rotation curve, thus causing 
a central depression in the integrated profile.  This, in turn, biases 
linewidths {\it high}.  The comparison of rotation curves for blue and 
red emission lines (e.g. Fig.~\ref{RCs}) shows no discernable effect due 
to opacity in the centers of these nearby, moderately inclined, disks (see 
also Prada 1996).  Extinction effects, even at H$\alpha$, are most noticeable 
for tilted disks with $i>84^\circ$ and only 21-cm fluxes should be trusted 
in such cases (Courteau \& Faber 1988).  While our null test is reassuring 
for local TF studies, extinction effects should be revisited in TF studies 
of distant galaxies. 

\section{TF Applications at High-Redshift}

The greatest challenge to comparing local TFRs to distant galaxies 
is the matching of galaxy families.  The nearby TF calibrators should 
be direct descendants of the more distant targets, or else 
structural/dynamical differences will be meaningless.  
Contrary to most local TF calibrations which are often heavily pruned 
(e.g. for cosmic flow studies), the range of Hubble types for distant 
galaxies is poorly controlled -- in large part due to resolution and 
cosmological dimming effects.

A proper TF calibration sample must include a broad and unbiased range 
of morphologies to sample the general emission-line galaxy population 
-- barred, starbursting, irregular, and all other types of spirals -- 
as might be seen at high redshift.  
Only the ``kitchen sink'' calibrations (with all spiral types and 
morphologies), such as those of Barton et~al. (2001) or Kannappan et~al. 
(2002), should 
be used in cosmological TF studies (provided the calibration uses
unambiguous, high-quality, magnitudes and linewidths, and is free 
of Hubble flow distortions).  
The intrinsic scatter of ``all-inclusive'' TF calibrations, 
$\sigma_{TF} \gtsim 0\famag5$, 
is a full 0\famag1-0\famag2 higher than values often quoted in studies 
of TF evolution and cosmic flows.  Standardization of calibration
and analysis techniques is also quite important, as scatter depends 
strongly on techniques.  Renewed interest has also been given in 
B-band TF calibrations which match the targets' rest-frame luminosities. 
These calibrations have highest dispersions and are most sensitive to 
dust extinction corrections, thus complicating the work of TF practioners. 

% The addition of early-type(redder) spirals in local samples 
% also steepens the TF slope (Kannappan et~al. 2002).  
% In particular, Sa galaxies deviate as a group from the nominal TFR, 
% as found by Rubin et~al. (1985), mainly due to the fact that they are 
% a redder population (than Sc's). 

\medskip

Other important effects to luminosities and linewidths of distant
galaxies which would mimic TF evolution include variations of the 
dust/gas and mass-to-light ratios, truncation of rotation curves 
due to interactions or $(1+z)^3$ surface brightness dimming, etc.  
 \ See CFS02 for more details. 

\medskip

Luminosity-linewidth studies of distant galaxies, though far from trivial, 
hold the promise of very exciting advancements in our understanding of the
evolution of galaxy populations.

\end{document}